\pdfoutput=1
\documentclass[conference, 9pt]{IEEEtran}
\IEEEoverridecommandlockouts

\usepackage{cite}
\usepackage{amsmath,amssymb,amsfonts}
\usepackage{algorithmic}
\usepackage{graphicx}
\usepackage{textcomp}
\usepackage{xcolor}
\def\BibTeX{{\rm B\kern-.05em{\sc i\kern-.025em b}\kern-.08em
    T\kern-.1667em\lower.7ex\hbox{E}\kern-.125emX}}

\newcommand{\vect}[1]{\ensuremath{\boldsymbol{\mathbf{#1}}}}
\newcommand{\loga}{\mathrm{log}}
\usepackage[acronym]{glossaries}
\include{glossaries}
\usepackage{nicefrac}
\usepackage{xcolor}
\usepackage{tikz}
\usepackage{stackengine}
\usetikzlibrary{positioning}
\usepackage[hidelinks]{hyperref}
\usetikzlibrary{fit}
\usepackage{pgfplots}
\pgfplotsset{compat=1.17} 
\usepgfplotslibrary{groupplots}
\usepackage{cleveref}
\usepackage{float}
\usepackage{placeins}
\usepackage{dblfloatfix}
\usepackage{siunitx}
\usepackage{multirow}
\usepackage{balance}
\usepackage{booktabs}

\begin{document}
\newacronym{ode}{ODE}{Ordinary Differential Equation}
\newacronym{mfa}{MFA}{Montreal Forced Aligner}
\newacronym{tts}{TTS}{Text-to-Speech}
\newacronym{vae}{VAE}{Variational Autoencoder}
\newacronym{elbo}{ELBO}{Evidence Lower Bound}
\newacronym{cnf}{CNF}{Continuous Normalizing Flow}
\newacronym{eer}{EER}{Equal Error Rate}
\newacronym{wer}{WER}{Word Error Rate}
\newacronym{fp}{FP}{False Positives}
\newacronym{fn}{FN}{False Negatives}
\newacronym{asr}{ASR}{Automatic Speech Recognition}
\newacronym{pvqd}{PVQD}{Perceptual Voice Qualities Database}
\newacronym{ve}{VE}{Voice Editing}
\newacronym[longplural=perceptual voice qualities]{pvq}{PVQ}{perceptual voice quality}
\newacronym{cape-v}{CAPE-V}{Consensus Auditory-Perceptual Evaluation of Voice}
\newacronym{mos}{MOS}{Mean Opinion Score}
\newacronym{smos}{SMOS}{Speaker Similarity MOS}
\newacronym{ccnf}{CCNF}{Conditional Continuous Normalizing Flow}

\title{Speech Synthesis along Perceptual Voice Quality Dimensions
}

\author{\IEEEauthorblockN{Frederik Rautenberg$^*$, Michael Kuhlmann$^*$, Fritz Seebauer$^\dagger$, Jana Wiechmann$^\dagger$, Petra Wagner$^\dagger$, Reinhold Haeb-Umbach$^*$}
\IEEEauthorblockA{
\textit{$^*$Paderborn University, Paderborn, Germany; $^\dagger$Bielefeld University, Bielefeld, Germany}
}
}

\maketitle

\begin{abstract}
While expressive speech synthesis or voice conversion systems mainly focus on controlling or manipulating abstract prosodic characteristics of speech, such as emotion or accent, we here address the control of perceptual voice qualities (PVQs) recognized by phonetic experts, which are speech properties at a lower level of abstraction. The ability to manipulate PVQs can be a valuable tool for teaching speech pathologists in training or voice actors. In this paper, we integrate a Conditional Continuous-Normalizing-Flow-based method into a Text-to-Speech system to modify perceptual voice attributes on a continuous scale. Unlike previous approaches, our system avoids direct manipulation of acoustic correlates and instead learns from examples.   We demonstrate the system's capability by manipulating four voice qualities: Roughness, breathiness, resonance and weight. Phonetic experts evaluated these modifications, both for seen and unseen speaker conditions. The results highlight both the system's strengths and areas for improvement.
\end{abstract}

\section{Introduction}
With the advent of \gls{tts} systems capable of generating natural speech, the scope of applications has significantly expanded. Many current works integrate prosody \cite{beck2022wavebender, omahony24_speechprosody, lameris2023prosody, skerry2018towards} or emotional \cite{triantafyllopoulos2023overview, chevi2024daisy} control to create more natural-sounding or personalized voices. While research into the synthesis or manipulation of speech properties at this high level of abstraction is matured, manipulating \glspl{pvq}, i.e., voice properties at an intermediate level of abstraction somewhere between low-level articulatory phonetic features and the above level, is widely unexplored. Examples of \glspl{pvq} are \textit{breathiness}, \textit{roughness} and \textit{hoarseness}\cite{singh2019profiling}. The ability to manipulate
\glspl{pvq} opens up new areas of applications for \gls{tts} systems. For instance, speech pathologists often lack suitable audio examples of speech disorders to teach new pathologists, which makes the explanation more challenging. A \gls{tts} system capable of manipulating \glspl{pvq} could be used to generate appropriate voices. Additionally, most speech signal processing systems are trained on typical voices, i.e. no speech disorders, leading to a gap in performance for individuals with atypical voices \cite{keller2004analysis}. Here, a \gls{tts} system could be used for data augmentation to address this gap.      

A recent work \cite{li2024articulatory} introduced the GTR-Voice dataset, which consists of recordings of a voice actor who manipulates his voice along three articulatory phonetic dimensions: \textit{Glottalization}, \textit{tenseness}, and \textit{resonance}. Utilizing expressive \gls{tts} systems, this work demonstrated the controllability of a synthesized voice along these dimensions. However, the modifications worked only for the voice of the actor producing the dataset.

The authors of \cite{netzorg2023permod} conditioned a latent diffusion model on a vector consisting of seven \glspl{pvq}, which are not observed independently across the dimensions. Two of these qualities are gender-related: \textit{Resonance} and \textit{weight}. The other, i.e., \textit{breathiness}, \textit{loudness}, \textit{pitch}, \textit{roughness}, and \textit{strain} are evaluated according to the \gls{cape-v} protocol \cite{kempster2009consensus}, which primarily aims to describe the severity of auditory-perceptual attributes associated with voice disorders. Note that the rating of \textit{pitch} reflects the perceptual deviation from what is considered normal. The approach showed acceptable performance in manipulating gendered \glspl{pvq} but had limitations for the other \glspl{pvq} as well as generating atypical voices.

Other approaches \cite{liuni2020angus, ruinskiy2008stochastic, verma2005introducing} focus on manipulating the acoustic correlates of the \glspl{pvq}. These studies have shown that \textit{jitter} and \textit{shimmer} are acoustic correlates of perceived \textit{roughness} and \textit{hoarseness}. In further experiments, they adjusted these acoustic correlates and shaped the listener's perception of \textit{roughness}, demonstrating a direct link between these acoustic features and the \gls{pvq}. However, while these modifications may lead to a change in perception, they do not fully capture all aspects of the \glspl{pvq}. Relying on such systems to teach speech pathologists may result in incomplete or potentially misleading explanations. 

In this work, we tackle the challenge of manipulating
gendered and non-gendered \glspl{pvq}, specifically: \textit{Breathiness}, \textit{roughness}, \textit{resonance}, and \textit{weight}. Given the largely unexplored nature of this field, we encounter several significant challenges. One of the primary issues is that the \glspl{pvq} considered in this study are not independent across dimensions, resulting in a lack of parallel data. Additionally, there are very few datasets that include annotated \glspl{pvq} and atypical voices. The available datasets are small, with even smaller subsets of atypical voices, and their annotations are often subject to high variability, as noted in \cite{netzorg2024towards}. Despite these challenges, we demonstrate that our system achieves acceptable performance in modifying \glspl{pvq} without directly altering acoustic correlates. The proposed system works for seen and unseen speakers during training. 

Normalizing flows \cite{rezende2015variational, kobyzev2020normalizing} showed their potential in manipulating global voice attributes as well as style continuously \cite{anastassiou2024voiceshop}. Motivated by this success, we use them in a neural \gls{tts} system to manipulate \glspl{pvq}. We use YourTTS \cite{casanova2022yourtts} as the speech synthesis system, which takes a text and a manipulated speaker representation from a Conditioned Continuous Normalizing Flow as input. The normalizing flow is conditioned by an 8-dimensional attribute vector containing the same \glspl{pvq} as in \cite{netzorg2023permod, netzorg2024towards}, plus the speaking rate. 

Manipulation performance was evaluated by adjusting the attribute vector along the axis of the selected \gls{pvq}. 
We demonstrate the system's  capabilities\footnote{Audio examples: \href{https://groups.uni-paderborn.de/nt/icassp_2025_deep_phonetics/norm_flow.html}{go.upb.de/ICASSP2025}} on the four 
\glspl{pvq} mentioned above.

As an approach for objective evaluation, we used pre-trained random forest regressors~\cite{netzorg2024towards} to quantify \glspl{pvq} and evaluated speaker embedding similarity after manipulation.

Additionally, we asked phonetic experts to evaluate manipulations of two \glspl{pvq}: \textit{Roughness} and \textit{breathiness}. Only two were selected due to the time-intensive nature of the evaluation process. These qualities were chosen because \cite{netzorg2024towards} demonstrated both high prediction performance of the regressors and lower deviation in expert ratings compared to other \glspl{pvq}. The experts rated the perceived severity of the \glspl{pvq} following the CAPE-V protocol  \cite{kempster2009consensus}, as well as assessing speaker similarity and audio quality. The results indicate that audio quality remains largely unaffected by the manipulations, and the predicted and perceived \gls{pvq} severity of the synthesized voice adjusts according to the conditioning. However, the findings also reveal that \textit{breathiness} can be manipulated more effectively than \textit{roughness} and that strong modifications reduce speaker similarity. 

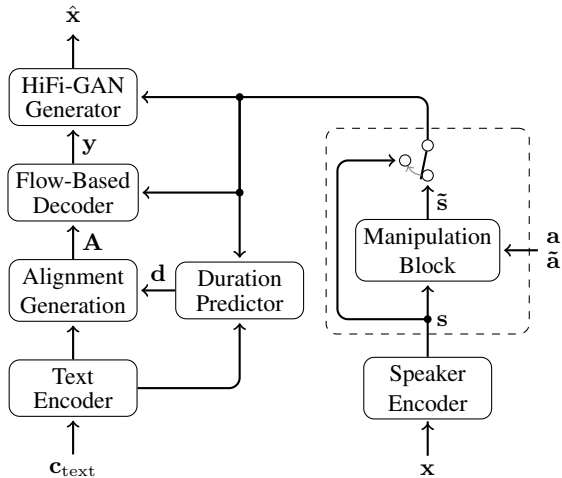
\begin{figure}[!b]
    \centering
    \tikzstyle{point} = [draw, fill=white, circle, radius=1cm, scale=0.5]
\tikzstyle{smallpoint} = [draw, fill=black, circle, radius=1cm, scale=0.3]
\begin{tikzpicture}[auto, node distance=0.5cm, 		
block/.style={
    rectangle,
    draw,
    text centered,
    minimum width=1.7cm, %
    minimum height=0.6cm, %
    rounded corners,
},
	]
	\node [] (input_text) {$\vect{c}_{\mathrm{text}}$};

    \node[block, rounded corners, above = of input_text] (TE) {\stackanchor{Text}{Encoder}};
    \node[block, rounded corners,  right = 3cm of TE] (SE) {\stackanchor{Speaker}{Encoder}};

    \node[block, rounded corners, above = 1cm of SE] (ode) {\stackanchor{Manipulation}{Block}};

    \node at (input_text.east -| SE.south) (speech_signal){$\vect{x}$};

    \node[block,  rounded corners, above = of TE](AG){\stackanchor{Alignment}{Generation}};
    \node[block, rounded corners, right = of AG] (DP) {\stackanchor{Duration}{Predictor}};
    \node[block,  rounded corners, above = of AG](FB){\stackanchor{Flow-Based}{Decoder}};
    \node[block,  rounded corners, above = of FB](hifi){\stackanchor{HiFi-GAN}{Generator}};

    \node[above=of hifi](speech_signal_output){$\hat{\vect{x}}$};

    \draw[->, thick] (input_text) -- ([shift=({0, -0.05})] TE.south);
    \draw[->, thick, rounded corners] (TE) -| ([shift=({0,-0.05})] DP.south);
    \draw[->, thick] (TE) -- ([shift=({0,-0.05})] AG.south);
    \draw[->, thick] (AG) -- node[right] {$\vect{A}$} ([shift=({0,-0.05})] FB.south);
    \draw[->, thick] (FB) -- node[right] {$\vect{y}$} ([shift=({0,-0.05})] hifi.south); 
    \draw[->, thick] (SE) -- node[right] {$\vect{s}$}([shift=({0,-0.05})] ode.south); 
    \draw[->, thick] (DP.west) -- node[above] {$\vect{d}$} ([shift=({ 0.05,0})] AG.east); 
    \draw[->, thick] (speech_signal.north) |- ([shift=({0, -0.05})] SE.south); 
    \draw[->, thick] (hifi.north) -- ([shift=({0, -0.05})] speech_signal_output.south); 

    \node[above = 0.5cm of ode.north, point] (point1) {};
    \node[point] at ($(point1) + (-0.3cm,0.2cm)$) (point2) {};
    \node[point] at ($(point1) + (0cm,0.4cm)$) (point3) {};
    \draw[-, thick] ([yshift=-0.05cm, xshift=-0.4] point1.west) -- (point3);
    \draw[->, color=gray] (point1) to [bend left] (point2);

    \draw[->, thick, rounded corners] ($(SE.north) + (0, 0.5cm)$)  -- ++(-1.2cm, 0.cm) |- ([shift=({-0.05, 0})] point2.west);
    \draw[->, thick] (ode.north) -- node[right] {$\vect{\tilde{s}}$}  ([shift=({0, -0.05})] point1.south);
    \draw[->, thick] (DP.north |-hifi.east)  -- ([shift=({0, 0.05})] DP.north);
    \draw[->, thick, rounded corners] (point3.north) |- ([shift=({0.05, 0})] hifi.east);
    \draw[->, thick] (DP.north |-hifi.east)  |- ([shift=({0.05, 0})]FB.east);

    \node[right = of ode](attribut){\stackanchor{$\vect{a}$}{$\vect{\tilde{a}}$}};
    \draw[->, thick] (attribut) -- ([shift=({0.05, 0})] ode.east);

    \node[smallpoint] at (DP.north |-hifi.east){};
    \node[smallpoint] at (DP.north |-FB.east){};
    \node[smallpoint] at ($(SE.north) + (0, 0.5cm)$) (sep) {};
    \begin{scope}
        \node[fit=(sep)(point3), draw, inner sep=0.15cm, rectangle, rounded corners, dashed, minimum width=2.7cm] {};
    \end{scope}

\end{tikzpicture}
    \caption{\gls{tts} system during inference \cite{casanova2022yourtts}. A speaker manipulation block is added to the system (dashed block). $\vect{a}$ are the estimated attributes from the input utterance $\vect{x}$ and $\vect{\tilde{a}}$ the desired ones of $\hat{\vect{x}}$.}
    \label{fig:tts}
\end{figure}

\section{Conditional Continuous Normalizing Flows}
\label{Sec:cnf}
Our voice modification system is built on the framework of Normalizing Flows \cite{rezende2015variational}. Below, we briefly review the concept of \glspl{ccnf} \cite{anastassiou2024voiceshop, abdal2021styleflow}.

Let $\vect{s}$ represent a random variable sampled from the distribution $p_{\vect{S}}(\vect{s} | \vect{a} ;\vect{\theta})$, which is parameterized by $\vect{\theta}$ and conditioned on an attribute vector $\vect{a}$. The objective is to sample from the modified distribution $p_{\vect{S}}(\vect{s} | \vect{\tilde{a}};\vect{\theta})$, where $\vect{\tilde{a}}$ denotes a manipulated attribute vector. This is accomplished by training a conditional generative model $p_{\vect{S}}(\vect{s} | \vect{a} ;\vect{\theta})$ and substituting $\vect{a}$ with $\vect{\tilde{a}}$ during the sampling process. The first step involves solving the initial value problem for the ordinary differential equation (ODE) \cite{chen2018neural, grathwohl2018ffjord}
\begin{equation}
    \label{Eq:transformation_normalizing_flow}
    \begin{aligned}
        \vect{z}(t_0) =  \vect{z}(t_1) + \int_{t_1}^{t_0} f(\vect{z}(t), t, \vect{a}, \vect{\theta}) \mathrm{d}t \, ,
    \end{aligned}
\end{equation}
with the initial condition $\vect{s} = \vect{z}(t_1)$ to obtain $\vect{z}(t_0) \sim \mathcal{N}(\mathbf{0}, \mathbf{I})$.
The function $f(\cdot)$ is parameterized as a neural network with parameters $\vect{\theta}$ and learned from the dataset  $\mathcal{S} = \left\{ \vect{s}_{n} \right\}_{n=1}^{N}$ by maximizing the log-likelihood \cite{anastassiou2024voiceshop, abdal2021styleflow} 
\begin{equation}
    \begin{aligned}
            l(\vect{\theta}) &=  \sum_{n=1}^N \loga \, p_{\vect{S}}(\vect{s}_n|\vect{a}_n;\vect{\theta})\\ 
        &= \sum_{n=1}^N \loga \, p_{\vect{Z}_0} \left(\vect{z}_{n}(t_0) \right) -  \Delta_{\loga p}(\vect{\theta}) \, ,
    \end{aligned}
    \label{eq:log_likelihood_cont}
\end{equation}
where i.i.d. samples are assumed and 
\begin{equation}
    \label{eq:second_ode}
    \begin{aligned}
        \Delta_{\loga p}(\vect{\theta}) =& \loga \, p_{\vect{Z}_1} \left(\vect{z}_{n}(t_1)\right) -  \loga \, p_{\vect{S}}(\vect{s}_n | \vect{a}_{n};\vect{\theta})  \\
        &- \int_{t_1}^{t_0}\mathrm{tr}\left( \frac{\mathrm{d} f(\vect{z}(t), t, \vect{a}_n, \vect{\theta})}{\mathrm{d}\vect{z}(t)} \right) \mathrm{d}t \, ,
    \end{aligned}
\end{equation}
which is also an initial value problem with the initial condition $\loga \, p_{\vect{Z}_1} \left(\vect{z}_{n}(t_1)\right) - \loga~p_{\vect{S}}(\vect{s}_n | \vect{a}_{n};\vect{\theta}) =0$. After training, a new sample $\tilde{\vect{s}}$ conditioned by a new attribute vector $\vect{\tilde{a}}$ can be generated with
\begin{equation}
    \begin{aligned}
        \tilde{\vect{s}} = \tilde{\vect{z}}(t_1) =  \vect{z}(t_0) + \int_{t_0}^{t_1} f(\vect{z}(t), t, \vect{\tilde{a}}, \vect{\theta}) \mathrm{d}t \, ,
    \end{aligned}
\end{equation}
where $\vect{z}(t_0)$ can be obtained either by sampling from a standard normal distribution, $\vect{z}(t_0) \sim \mathcal{N}(\mathbf{0}, \mathbf{I})$, or by applying the transformation given in \Cref{Eq:transformation_normalizing_flow} to a given sample $\vect{s}$.
\section{Voice-Editing Supporting Text-to-Speech Synthesis}
\subsection{Speech Synthesis System}
We utilize YourTTS \cite{casanova2022yourtts} as the speech synthesis system. This model is an adaptation from the VITS model \cite{kim2021conditional}, in which the training objective is the maximization of the \gls{elbo} of the log-likelihood function 
\begin{equation}
    \begin{aligned}
        \loga \, p_{\vect{X}}(\vect{x} | \vect{c}; \vect{\psi}) \geq  \mathbb{E}_{q_{\vect{Y}}(\mathbf{y}|\vect{x}; \boldsymbol{\phi})} \left[ \vphantom{\frac{test}{test}} \right.& \loga \,p_{\vect{X}}(\vect{x} | \vect{y}; \vect{\psi})   \\ &- \left. \loga\frac{q_{\vect{Y}}(\mathbf{y}|\vect{x}; \vect{\phi})}{p_{\vect{Y}}(\mathbf{y}|\vect{c}; \vect{\psi})} \right] \, ,   
    \end{aligned}
\end{equation}
with the learnable parameters $\vect{\psi}$ and $\vect{\phi}$, latent random variable $\vect{y}$ and the condition $\vect{c}= \left[ \vect{c}_{\mathrm{text}}, \vect{A}, \vect{s}\right]$, which is a combination of the text embedding $\vect{c}_{\mathrm{text}}$, the alignment $\vect{A}$,  and the speaker embedding $\vect{s}$. 
\begin{table}[!t]
\caption{Performance of the trained random forest regressors on the PVQD for each \gls{pvq} and their description}
\begin{center}
\begin{tabular}{lc>{\arraybackslash}m{4.5cm}} %
\toprule[1pt]
\gls{pvq} & RMSE $\downarrow$ & Description \cite{netzorg2023permod} \\
\midrule
Resonance & 17.72 & Sound quality of the size of the vocal tract \\
\hline
Weight & 16.48 & Sound quality of the vocal fold vibratory mass \\
\hline
Strain & 10.66 & Perception of excessive vocal effort (hyperfunction) \\
\hline
Loudness & 9.48 & Deviation in loudness \\
\hline
Roughness & 7.70 & Perceived irregularity in the voicing source \\
\hline
Breathiness & 9.06 & Audible air escape in the voice \\
\hline
Pitch & 7.88 & Deviation in pitch \\
\bottomrule[1pt]
\end{tabular}
\label{tab:rmse_pvq}
\end{center}
\vspace{-0.4cm}
\end{table}
\begin{table*}[!t]
\newcommand{\meanstd}[2]{\num[round-mode=places, round-precision=0]{#1} {\footnotesize \textcolor{gray}{$\pm$\,\num[round-mode=places, round-precision=0]{#2}}}}
\caption{Predicted PVQ from random forest regressors and EER for seen and unseen speakers of manipulated voices. Expected perceived severities are low (0), medium (50), and high (100).}
\begin{center}
\setlength{\tabcolsep}{3.5pt}
\begin{tabular}{cccccccccccccc} 
\toprule
 & Set & \multicolumn{3}{c}{Roughness}  &  \multicolumn{3}{c}{Breathiness}  &  \multicolumn{3}{c}{Resonance}  &  \multicolumn{3}{c}{Weight} \\
 \cmidrule(l){3-5} \cmidrule(l){6-8} \cmidrule(l){9-11} \cmidrule(l){12-14}
 & & Low & Med. & High & Low & Med. & High & Low & Med. & High & Low & Med. & High \\ 
 \midrule 
 \multirow{2}{*}{Pred. PVQ} & seen & \meanstd{14.8}{3} & \meanstd{22.0}{3} & \meanstd{28.0}{4} & \meanstd{11.3}{2} & \meanstd{19.4}{4} & \meanstd{35.3}{6} & \meanstd{31.4}{7} & \meanstd{39.8}{10} & \meanstd{49.0}{7} & \meanstd{34.5}{5} & \meanstd{48.9}{10} & \meanstd{63.8}{11} \\
  & unseen & \meanstd{14.5}{3} & \meanstd{21.1}{3} & \meanstd{27.6}{4} & \meanstd{11.1}{2} & \meanstd{19.4}{3} & \meanstd{35.3}{6} & \meanstd{31.7}{7} & \meanstd{39.8}{9} & \meanstd{48.9}{6} & \meanstd{34.9}{4} & \meanstd{48.3}{7} & \meanstd{63.9}{10}\\
 \midrule 
 \multirow{2}{*}{EER $[\%] \downarrow$} & seen & 1.4 & 3.8 & 10.2 & 1.2 & 2.3 & 8.0&8.8&1.4&22.2&29.6&1.7&19.4\\
  & unseen & 2.2 & 3.6 & 10.1 & 1.9 & 2.5 & 6.3&8.0&1.5&23.3&30.8&2.1&16.3\\
  \bottomrule
\end{tabular}
\label{tab:eer_pvq}
\end{center}
\vspace{-0.45cm}
\end{table*}
We made some modifications to the original architecture proposed in \cite{casanova2022yourtts}. Specifically, we replaced the stochastic duration predictor with a deterministic one and substituted the speaker encoder with a pre-trained d-vector model \cite{cord2023frame}. The d-vector model is a ResNet34 model that computes 256-dimensional speaker embeddings from 80-dimensional log-mel spectrograms \cite{cord2023frame}. \Cref{fig:tts} shows the overall system during inference, where $\hat{\vect{x}}$ is the synthesized speech signal containing the content of $\vect{c}_{\mathrm{text}}$, the speaker characteristics of the speech signal $\vect{x}$ up to the modified \glspl{pvq} in $\vect{\tilde{a}}$. We trained the \gls{tts} system on LibriTTS-R \cite{koizumi2023libritts}, an enhanced version of LibriTTS \cite{zen2019libritts} with improved sound quality. The dataset comprises 585 hours of speech data at $24\,\text{kHz}$ from 2,456 speakers.

\subsection{Flow-Based Parametric Speech Modification}
\label{Sec:attribute_manipulation}
We extract speaker embeddings $\vect{s}_n$ from the pre-trained speaker encoder and assume that these are samples from the distribution $p_{\vect{S}}(\vect{s}_n | \vect{x_n}, \vect{a}_n;\vect{\theta})$, where $\vect{x}_n$ is the input utterance and $\vect{a}_n$ is the attribute vector containing \glspl{pvq} and the speaking rate. We further assume that $\vect{x_n}$ is implicitly conditioned on $\vect{a}_n$, eliminating the need for additional input to the speaker encoder. Our goal is to adjust the attributes $\vect{a}_n$ of the speaker embedding $\vect{s}_n$, such that the transformed embedding $\vect{s}_n$ differs only in the updated conditioning $\tilde{\vect{a}}_n$. This modification is achieved using a \gls{ccnf}, as described in \Cref{{Sec:cnf}}.

We used seven \glspl{pvq} from the extended Perceptual Voice Qualities Database (PVQD+)~\cite{netzorg2024towards}. This dataset is an addition to \gls{pvqd} \cite{walden2022perceptual}, where 3 expert clinicians followed the \gls{cape-v} \cite{kempster2009consensus} evaluation protocol to rate the perceived severity of utterances from participants with typical and atypical voices. This score starts at 0 which means no severity (healthy) and goes up to 100 which means full severity (dysarthric). The rated \glspl{pvq} are \textit{strain}, \textit{roughness}, \textit{breathiness}, \textit{pitch}, and \textit{loudness}, with the authors of PVQD+ \cite{netzorg2024towards} adding labels for \textit{weight} and \textit{resonance}. This dataset consists of 296 annotated audio files including atypical voices. To overcome the limitation of the small dataset, we conducted a pseudo labeling of these \glspl{pvq} on the LibriTTS-R dataset. For the labeling, we followed \cite{netzorg2024towards} and trained random forest regressors for each \gls{pvq} in PVQD+ and used the same train test splits. Random forests were fitted on time-averaged hidden representations of the 6th layer of a HuBERT Large~\cite{hsu2021hubert} model, 
these features proved the best predictors of speaker identity~\cite{chen2021wavlm}, and achieved an average test RMSE of $11.28$ which is in line with~\cite{netzorg2024towards}. The RMSE for each \gls{pvq} is shown in \Cref{tab:rmse_pvq}. These fitted regressors were used to pseudo-label the LibriTTS-R dataset.
For the full training data (PVQD and pseudo-labeled LibriTTS-R), we observed that the range of non-gendered \glspl{pvq} is mostly distributed in the first half of the CAPE-V scale. The severity values for the \glspl{pvq} are distributed with $20\textcolor{gray}{\,\pm\,17}$ for \textit{roughness}, $19\textcolor{gray}{\,\pm\,20}$ for \textit{breathiness}.
We attribute the limited amount of observed atypical voices to this data skew. 
Despite this bias, the model was able to synthesize \glspl{pvq} with sufficient intensities~(see \Cref{results}).
For \textit{weight} and \textit{resonance}, we get prediction scores of 54$\textcolor{gray}{\,\pm\,18}$ and 45$\textcolor{gray}{\,\pm\,15}$, respectively.

We observed a change in speaking rate after manipulation when it wasn't included in the attribute vector. Therefore, we calculated the speaking rate by extracting phonemes and their durations using the \gls{mfa} \cite{mcauliffe17_interspeech}, then dividing the number of phonemes by the total duration of the voiced segments.

\begin{figure}[!b]
    \centering
    \vspace{-0.4cm}
    \begin{tikzpicture}[auto, node distance=0.5cm, 		
block/.style={
    rectangle,
    draw,
    text centered,	
},
]
\node [] (input) {$\vect{s} = \vect{z}(t_1)$};
\node[block, rounded corners, below = 0.4cm of input](ODE_1){$\mathrm{ODESolve} \left(\cdot\right)$};
\node [right = of ODE_1] (input_2) {$\vect{a}$};

\node [below = 0.4cm of ODE_1] (s_0) {$\mathbf{z}(t_0)$};
\node[block, rounded corners, below = 0.4cm of s_0](ODE_2){$\mathrm{ODESolve} \left(\cdot\right)$};
\node [below = 0.4cm of ODE_2] (Output) {$\vect{s}_{\mathrm{m}} = \vect{z}_{\mathrm{m}}(t_1)$};

\node [right = of ODE_2] (input_man) {$\vect{a}_{\mathrm{m}}$};

\draw [->, thick] (input) -- ([shift=({0, 0.05})] ODE_1.north);
\draw [->, thick] (ODE_1) -- (s_0);
\draw [->, thick] (s_0) -- ([shift=({0, 0.05})] ODE_2.north);
\draw [->, thick] (ODE_2) -- (Output);

\draw [->, thick] (input_2) -- ([shift=({0.05, 0})] ODE_1.east);
\draw [->, thick] (input_man) -- ([shift=({0.05, 0})] ODE_2.east);

\node at ($(ODE_1.north) + (-1.5cm,0.2cm)$) (int_point_1) {$t_1$};
\node at ($(ODE_1.south) + (-1.5cm,-0.2cm)$) (int_point_2) {$t_0$};
\draw [->, thick] (int_point_1.south) -- (int_point_2.north);

\node at ($(ODE_2.north) + (-1.5cm,0.2cm)$) (int_point_3) {$t_0$};
\node at ($(ODE_2.south) + (-1.5cm,-0.2cm)$) (int_point_4) {$t_1$};
\draw [->, thick] (int_point_3.south) -- (int_point_4.north);

\end{tikzpicture}
    \caption{Speaker embedding manipulation block, where $\mathrm{ODESolve} \left(\cdot\right)$ gives a solution for the initial value problem}
    \label{fig:ode_manipulation_block}
\end{figure}
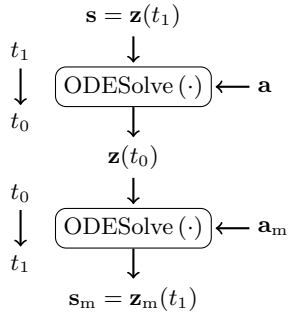

The \gls{ccnf} was trained similarly to the methods proposed in \cite{anastassiou2024voiceshop,abdal2021styleflow}, by maximizing the log-likelihood of $p_{\vect{S}}(\vect{s}_n | \vect{x}_n, \vect{a}_{n};\vect{\theta})$ as explained in \Cref{Sec:cnf}. We trained a single \gls{ccnf} for all seven \glspl{pvq} and the speaking rate on the training set of LibriTTS-R and the complete \gls{pvqd}. All values in the attribute vector are normalized to the interval $[0,1]$ before being conditioned to the normalizing flow. For the speaking rate, the maximum speaking rate of the train set is used as a reference. The \gls{ode} solver from \cite{chen2018neural} was employed with default tolerance parameters, and the trace was estimated using Hutchinson’s trace estimator as proposed in \cite{grathwohl2018ffjord}. The function $f(\cdot)$, see \Cref{Eq:transformation_normalizing_flow}, was modeled with one \gls{ccnf} block as introduced in \cite{abdal2021styleflow}, with a hidden size of $512$. We observed overfitting when using a larger size. Training was conducted with a batch size of 200 and an initial learning rate of $10^{-4}$. The learning rate was updated every 100 epochs by multiplying it with a decay factor of $\gamma = 0.98$. 
After training, manipulation was performed as illustrated in  \Cref{fig:ode_manipulation_block}. The speaker embedding was first transformed to $\vect{z}_0$ using the speaker attributes $\vect{a}_n$ and then back using the desired attributes $\vect{\tilde{a}}_{n}$. The modified embedding was then used to synthesize the speech signal $\hat{\vect{x}}$ via the \gls{tts} system, see \Cref{fig:tts}.

\begin{table*}[!t]
\newcommand{\meanstd}[2]{\num[round-mode=places, round-precision=1]{#1} {\footnotesize \textcolor{gray}{$\pm$\,\num[round-mode=places, round-precision=1]{#2}}}}
\caption{PVQ across different manipulation dimensions and severities for seen and unseen speakers, including the evaluations on the original speech. Breathiness severities are low (0), medium (50), and high (100) and roughness severities are low (0), medium (100), and high (200). The table presents averaged Mean Opinion Score (MOS) ratings on a 5-point scale (1 = Bad, 5 = Excellent) and Similarity Mean Opinion Score (SMOS) ratings on a 4-point scale (1 = Different speakers, 4 = Same speaker).
}
\begin{center}
\begin{tabular}{c@{\hspace{0.1cm}}cccccccc} %
\toprule
 & \multirow{2}{*}{Set} & \multirow{2}{*}{\stackanchor{Original}{Recording}} & \multicolumn{3}{c}{Roughness Manipulation}  &  \multicolumn{3}{c}{Breathiness Manipulation}\\
 \cmidrule(l){4-6} \cmidrule(l){7-9}
 & & & Low & Med. & High & Low & Med. & High \\ 
 \midrule 
\multirow{2}{*}{Perc. Roughness} & seen &  \meanstd{29.90}{24.25} & \meanstd{31.03}{25.0} & \meanstd{39.70}{30.3} & \meanstd{71.63}{22.6} & \meanstd{32.43}{24.5} & \meanstd{31.40}{23.0} & \meanstd{37.80}{28.4}\\
  & unseen &  \meanstd{29.88}{25.68}  & \meanstd{33.37}{28.4} & \meanstd{39.13}{32.4} & \meanstd{64.70}{29.5} & \meanstd{35.77}{26.2} & \meanstd{30.87}{28.4} & \meanstd{28.40}{27.2}\\
\midrule
\multirow{2}{*}{Perc. Breathiness} & seen & \meanstd{23.68}{22.43} & \meanstd{23.97}{19.7} & \meanstd{35.43}{27.4} & \meanstd{50.00}{30.7} & \meanstd{23.70}{21.7} & \meanstd{45.87}{32.9} & \meanstd{70.03}{28.8}\\
  & unseen & \meanstd{23.81}{24.12} & \meanstd{23.90}{23.0} & \meanstd{41.20}{24.1} & \meanstd{51.83}{31.2} & \meanstd{26.16}{26.7} & \meanstd{52.97}{33.6} & \meanstd{79.20}{27.0}\\
 \midrule 
 \midrule
 \multirow{2}{*}{MOS $\uparrow$ (1-5)} & seen & \meanstd{4.10}{1.0} & \meanstd{3.73}{1.3} & \meanstd{3.67}{1.2} & \meanstd{3.10}{1.5} & \meanstd{3.83}{1.2} & \meanstd{3.83}{1.2} & \meanstd{3.50}{1.2}\\
  & unseen & \meanstd{4.30}{1.0} & \meanstd{3.70}{1.3} & \meanstd{3.53}{1.2} & \meanstd{3.03}{1.5} & \meanstd{3.67}{1.5} & \meanstd{3.83}{1.3} & \meanstd{3.53}{1.4}\\
\midrule
\multirow{2}{*}{SMOS $\uparrow$ (1-4)} & seen & \meanstd{3.40}{1.0} & \meanstd{3.30}{1.0} & \meanstd{2.60}{1.2} & \meanstd{2.43}{0.9} & \meanstd{3.20}{1.0} & \meanstd{2.63}{1.0} & \meanstd{2.10}{1.0}\\
  & unseen & \meanstd{3.58}{0.8} & \meanstd{3.40}{0.9} & \meanstd{2.54}{1.1} & \meanstd{2.03}{1.2} & \meanstd{3.33}{1.0} & \meanstd{2.53}{1.1} & \meanstd{2.33}{0.9}\\
\bottomrule
\end{tabular}
\label{tab:results_perceived_pvq}
\end{center}
\vspace{-0.5cm}
\end{table*}

\section{Results and Discussions}
\label{results}
All modifications are performed by first extracting the attribute vector $\vect{a}$ of an utterance with the trained regressors and then manipulating a single \gls{pvq} along its axis while keeping the others fixed. We evaluated this modification process with four \glspl{pvq}: \textit{Breathiness}, \textit{roughness}, \textit{resonance} and \textit{weight}. Speaker embeddings were sampled and conditioned on varying strength levels: 0, 50, and 100, which we interpret as low, medium, and high. Due to the limitations of extreme cases of \textit{roughness} in the training data, the model struggled to map modifications across the full strength spectrum accurately. As a result, for this \gls{pvq}, the modification range was adjusted to 0, 100, and 200.

Objective evaluation included \gls{eer} to measure speaker similarity and the random forest regressors used for pseudo-labeling to evaluate the severity of the modified \glspl{pvq}.
We evaluated 200 seen male and female speakers from the training set and 39 (20 male, 19 female) unseen speakers from the test set of LibriTTS-R, using up to 10 utterances per speaker. Note that the dataset lacks extreme speakers, limiting the non-gendered \gls{pvq} manipulations from low to high.
The \gls{eer} was calculated by extracting speaker embedding $\hat{\vect{s}}$ from the synthesized utterance $\hat{\vect{x}}$ using the same speaker encoder employed in the \gls{tts} system.
Speaker embeddings were also extracted from the training set, and positive and negative pairs were constructed.
Cosine similarity was used to measure the similarity between embeddings.
When a manipulation was applied, the positive pair consisted of the manipulated embedding and its unmanipulated counterpart.

To evaluate the perceived change in voice quality, a subjective listening test was conducted.
15 raters were asked to evaluate 480 samples in total on a digital version of the 100-point scales employed in \cite{kempster2009consensus}.
The presented scales were labeled \textit{roughness} and \textit{breathiness}, respectively.
All participants were phonetic experts and recruited by word of mouth. 
Each participant was presented with the manipulated samples in 3 different modification levels for \textit{roughness} and \textit{breathiness} as described above, plus an original recording of the speaker, shuffled in random order.
Each participant rated 32 samples, synthesized from 4 different speakers, stratified by gender and whether the speaker was seen in the training set.
Sample lengths ranged from $5 - 29\,\mathrm{s}$, with an average of $11\,\mathrm{s}$.
In order to ensure that the manipulation of the \glspl{pvq} did not adversely affect the synthesis quality, the raters were additionally asked to rate the overall perceived quality of each sample using \gls{mos}, with the standard ITU-R scale ranging from \textit{Bad} to \textit{Excellent} \cite{ITU808}.
As a final task, they were also given a speaker similarity test using \gls{smos} \cite{yi20_vccbc} with a random reference from the original speaker to examine whether the change in voice quality distorts the speaker perception.
The results for objective and subjective evaluation are given in~\Cref{tab:eer_pvq} and~\Cref{tab:results_perceived_pvq}, respectively.

\subsection{Manipulating Voice Quality Intensity}
The upper halves of the tables show the predicted and perceived \gls{pvq} scores.
While all values increase with increasing conditioning, the scores predicted by the random forest regressors are confined to a much smaller interval than the expert ratings.
We attribute the score skew in PVQD on which the regressors were trained to this discrepancy, demanding awareness when relying on regression models to judge voice quality.
For the perceived ratings, a paired Wilcoxon rank sum test \cite{bauer1972} was conducted between all conditions to confirm the statistical significance of the distributional shift.
In the \textit{breathiness} condition, all changes between manipulation levels proved to be significant ($p < 0.005$), except for the difference from the original recording to low manipulation, which constitutes pure re-synthesis as all manipulations are done on typical voices, i.e., low severity of the \glspl{pvq}.
In the \textit{roughness} condition, only the perceived changes from all levels to high modification proved statistically significant ($p < 0.005$).

Looking at the expert ratings, we observe that, indeed, an aggressive roughness manipulation was necessary to get a high score.
Simultaneously, perceived breathiness increases when increasing roughness, while manipulation of breathiness has a mixed effect on perceived roughness.
A recent study on the effect between rough and breathy voice qualities~\cite{park2022interactions} found that increased roughness has a stronger effect on perceived breathiness than increased breathiness on perceived roughness. Still, the effect in both directions is small. 
Rather than being an artifact of entanglement in the normalizing flow, this effect seems to happen naturally for these two voice qualities.
The slight downward shift in \gls{mos} did not prove to be statistically significant, except from the original reference to high manipulation in both \textit{roughness} and \textit{breathiness} conditions ($p < 0.005$).

\subsection{Effect on Speaker Similarity}
The lower halves of the tables present speaker similarity findings after manipulation.
As the severity of the \glspl{pvq} increases, the \gls{eer} also rises, suggesting that speaker identity preservation becomes more challenging. This is expected since modern speaker embedding models are optimized to distinguish small perceptual nuances between different voices.
A similar trend can be observed for the SMOS.
Here, we have significant degradation in speaker identification between all levels, except for the original reference and low manipulation ($p < 0.005$).
Remember that manipulations were performed on LibriTTS-R, so all speakers have typical voices characterized by low \textit{roughness} and \textit{breathiness} values.
Modifications with lower levels of these \glspl{pvq} result in a lower \gls{eer}, as the synthesized voices remain more similar to the original speaker.
The findings could point to the notion that voice quality is an integral part of the speaker's identity and cannot be perceived separately.
We will leave it to future work to verify this.

\subsection{Manipulation of Gendered \glspl{pvq}}
Lastly, we evaluated the modification of the two gendered \glspl{pvq} \textit{resonance} and \textit{weight}.
The objective evaluation results are shown in~\Cref{tab:eer_pvq}.
Given that the expert ratings for these \glspl{pvq} were distributed more uniformly across the full scale~\cite{netzorg2024towards}, we expect the regression scores to be more reliable.
Again, we observe a clear increasing trend with increasing conditioning, supporting our approach's success in not only manipulating gendered (as in \cite{netzorg2023permod}) but also non-gendered \glspl{pvq}. The impact we observe on speaker similarity is much greater than for the non-gendered \glspl{pvq}.
Note that \textit{weight} should be high for healthy
male voices and low for healthy female voices \cite{netzorg2024towards}, and vice versa for
\textit{resonance}.
Strong manipulation of these values alters perceived gender, spoofing the \gls{eer} to a greater extent.
This also explains the lower \gls{eer} for medium manipulations.

\section{Conclusion}
The experiments demonstrated that a \gls{tts} approach based on Conditional Continuous Normalizing Flows effectively manipulates \glspl{pvq}. Despite being trained on only a limited set of atypical voices, the model showed solid performance in modifying four \glspl{pvq} for both seen and unseen voices. Notably, changes in two of these \glspl{pvq} were convincing to phonetic experts. We hypothesize that a tool that can apply those modifications to a voice can be used by speech pathologists in teaching new pathologists. By generating speech samples with specific perceptual properties selectively altered, these tools can assist in illustrating or replicating particular perceptual voice categories. Our goal is to verify this hypothesis in future work.

\section*{Acknowledgement}
Funded by the Deutsche Forschungsgemeinschaft (DFG, German Research Foundation): TRR 318/1 2021- 438445824 and 446378607.

\newpage

\bibliographystyle{IEEEbib}
\bibliography{references}
\end{document}